\documentclass[aps,pra,twocolumn,reprint,superscriptaddress,nofootinbib]{revtex4-2}

\usepackage[svgnames,table,xcdraw]{xcolor}
\usepackage{amsmath}
\usepackage{microtype}
\usepackage{amssymb}
\usepackage{braket}
\usepackage{graphicx}
\usepackage{booktabs}
\usepackage{bm}
\usepackage{dsfont}
\usepackage{hyperref}
\usepackage[caption=false]{subfig}
\usepackage{amsthm}

\usepackage{xfrac}
\usepackage{enumitem}
\usepackage{orcidlink}
\usepackage{appendix}
\usepackage[ruled,vlined]{algorithm2e}

\hypersetup{colorlinks,linkcolor={DarkBlue},citecolor={DarkBlue},urlcolor={DarkBlue}}

\let\originalleft\left
                     \let\originalright\right
\renewcommand{\left}{\mathopen{}\mathclose\bgroup\originalleft}
\renewcommand{\right}{\aftergroup\egroup\originalright}

\usepackage{mathtools}

\DontPrintSemicolon
\SetKwInput{KwInput}{Input}
\SetKwInput{KwOutput}{Output}
\SetKwFor{ForEach}{foreach}{do}{}
\SetKwFor{ForLoop}{for}{do}{}

\usepackage{nag}
\usepackage{graphicx}
\usepackage{amsthm}
\usepackage{tabulary}
\usepackage{qcircuit}
\usepackage{mathtools}
\usepackage{bm}
\usepackage{braket}
\usepackage{datetime}
\usepackage{stackrel}
\usepackage{dsfont}
\usepackage{qcircuit}
\usepackage[english]{babel}
\usepackage{units}

\usepackage{tikz}
\usetikzlibrary{backgrounds,decorations.pathreplacing}

\usepackage{chngcntr}
\counterwithout{equation}{section}

\usepackage[normalem]{ulem}
\usepackage{slashed}
\usepackage{soul}

\usepackage{xcolor}
\usepackage{amssymb}
\usepackage{amsmath}
\usepackage{amsthm}
\usepackage{braket}
\usepackage{mathdots}

\usepackage{makeidx}
\makeindex

\usepackage{soul}
\usepackage{xargs}
\newcommandx{\cmnote}[2][1=]{\linespread{1.0}\todo[linecolor=red,backgroundcolor=red!25,bordercolor=red,#1]{#2}}

\let\underline\ul

\allowdisplaybreaks[4]

 \index{} \index{} \index{} \index{} \index{} \index{} \index{} \index{}
\makeatletter

\newcommand{\ringplus}{\mathbin{\text{\@ringplus}}}

\newcommand{\@ringplus}{%
  \ooalign{\hidewidth\raise1.3ex\hbox{\tiny$\circ$}\hidewidth\cr$\m@th+$\cr}%
}

\newcommand{\ringminus}{\mathbin{\text{\@ringminus}}}

\newcommand{\@ringminus}{%
  \ooalign{\hidewidth\raise0.9ex\hbox{\tiny$\circ$}\hidewidth\cr$\m@th-$\cr}%
}
\makeatother
 \index{} \index{} \index{} \index{} \index{} \index{} \index{} \index{}

\DeclareFontFamily{U}{wncy}{}
\DeclareFontShape{U}{wncy}{m}{n}{<->wncyr10}{}
\DeclareSymbolFont{mcy}{U}{wncy}{m}{n}
\DeclareMathSymbol{\Sh}{\mathord}{mcy}{"58}

 \index{} \index{} \index{} \index{} \index{} \index{} \index{} \index{}
 \index{} \index{} \index{} \index{} \index{} \index{} \index{} \index{}

 \index{} \index{} \index{} \index{} \index{} \index{} \index{} \index{}
 \index{} \index{} \index{} \index{} \index{} \index{} \index{} \index{}
\usepackage{graphicx}

\usepackage{graphicx}
\usepackage{multirow}
\usepackage{hhline}

\xyoption{color}
\xyoption{line}

\newcommandx*\bsbal[3][1=black, 3=->]{\ar @[#1]@{#3} [#2,0] \qw}

\newcommandx*\varbs[5][1=black, 3=\theta,4=0.5,5=->]{\ar @[#1]@{#5}^(#4){#3} [#2,0] \qw}

\newcommandx*\lblline[3][3=0.5]{\ar @{-}^(#3){#1} [#2,0]}
\newcommandx*\ctrlg[3][3=0.5]{ \raisebox{-3pt}{$\bullet$}  \ar @{-}^(#3){#1} [#2,0] \qw }
\newcommandx*\ctrlog[2]{\controlo \ar @{-}^{#1} [#2,0] \qw}
\newcommandx*\ctrlodash[1]{\controlo \ar @{-} [#1,0] \ar @[black]@{.} [0,-1]}

\begin{document}

    \title{%
        \texorpdfstring
        {Decoder Comparability Across Quantum Software Stacks: \\Repeated-Round Surface and Digitized-GKP Syndrome Replay}
        {Decoder Comparability Across Quantum Software Stacks: Repeated-Round Surface and Digitized-GKP Syndrome Replay}
    }

    \def \affGatech {College of Computing, Georgia Institute of Technology, Atlanta, GA 30332 USA}
    \def \vgg {Volkswagen AG, Berliner Ring 2, Wolfsburg 38440, Germany}
    \def \rwth {Department of Physics, RWTH Aachen, Germany}
        \def \dop {Department of Physics, Federal University of Juiz de Fora, Juiz de Fora, 36036-900, Brazil}
    \def \fujf {Department of Computational and Applied Mechanics, \\Federal University of Juiz de Fora, Juiz de Fora, 36036-900, Brazil}
    \def \kfup {Civil and Environmental Engineering Department, King Fahd University of Petroleum \\\& Minerals, Dhahran 31261, Saudi Arabia}
    \def \tubaf {Institute of Computer Science, Faculty of Mathematics and Computer Science, TU Bergakademie Freiberg, Bernhard-von-Cotta-Straße 2, D-09599 Freiberg, Germany}

    \author{Dennis Delali Kwesi Wayo \orcidlink{0000-0001-9980-6247}}
    \affiliation{\affGatech}
    \affiliation{\tubaf}

    \author{Chinonso Onah \orcidlink{0000-0002-6296-533X}}
    \affiliation{\vgg}
    \affiliation{\rwth}

    \author{Rodrigo Alves Dias \orcidlink{0000-0001-5638-4355}}
    \affiliation{\dop}
    
    \author{Leonardo Goliatt \,\orcidlink{0000-0002-2844-9470}}
    \affiliation{\fujf}

    \author{Zaher Mundher Yaseen \,\orcidlink{0000-}}
    \affiliation{\kfup}

    \author{Sven Groppe \,\orcidlink{0000-0001-5196-1117}}
    \affiliation{\tubaf}

    \date{\today}

    \begin{abstract}
        We present a contract-preserving, family-aware comparison of decoder behavior
        across four syndrome-generation stacks (PennyLane, Qiskit, Cirq, and a LiDMaS+
        reference) under a fixed replay interface. Request streams from repeated-round
        surface-code and digitized-GKP circuits are replayed through BP, MWPM, and UF
        with matched controls. The unified matrix spans 24 cells and achieves line-level
        integrity: 24\,000 request lines, 24\,000 response lines, response ratio
        \(=1.0\) in every cell, zero parse failures, and zero decoder-name mismatches.
        Fifteen warning-no-syndrome events occur only in GKP-Cirq rows. Within-family
        ordering is stable in both families (\(\mathrm{BP}<\mathrm{MWPM}<\mathrm{UF}\));
        source-averaged flip counts are 2.435, 4.768, and 5.870 for surface and
        1.595, 2.908, and 3.681 for GKP. Relative to MWPM, BP reduces mean intervention
        volume by \(48.9\%\) in surface and \(45.1\%\) in GKP. Source-vs-reference
        effects are family dependent, with larger coherent shifts in GKP. Source-bootstrap
        ranks remain unchanged, and hidden-truth sidecars add an outer-code
        logical-parity error-rate check in which UF has the largest source-mean rate
        in both families. The comparison is stack-aware, contract-verified, and
        avoids raw cross-family threshold-equivalence claims.
    \end{abstract}

    \maketitle

    \section{Introduction}
    Fault-tolerant quantum computing depends not only on error-correcting codes, but
    also on comparable implementations across code family, noise model,
    syndrome-generation path, and decoder policy. Surface codes are the main
    discrete-variable reference architecture because they combine local checks,
    scalable decoding, and well-studied thresholds
    \cite{surfacecode1998,fowler2009highthreshold,fowler2012proofmatching,wang2011surfaceover1,
        bravyi2024ldpcmemory,dua2024clifford}.
    Within that landscape, decoder families such as minimum-weight perfect matching
    (MWPM), union-find (UF), and belief-propagation-like or heuristic variants
    offer distinct tradeoffs in correction quality, latency, and implementation
    complexity
    \cite{duclos2010fast,unionfind2017,pymatching2021,higgott2023improved,
        skoric2023parallel,tan2023scalable}. Code deformations and bias-aware
    constructions, including XZZX-like families, further show that decoder
    performance depends on syndrome statistics as well as nominal code distance
    \cite{bonilla2021xzzx,tuckett2018ultrahigh,tuckett2020faulttolerant,
        lee2021rectangular,darmawan2021xzzxkerrcat,xu2023tailored}. Decoder
    comparisons are therefore only as reliable as the syndrome streams being
    compared.

    Recent hardware-scale demonstrations reinforce this point. Large surface-code
    experiments have shown suppressible logical error rates under scaling and
    below-threshold operation in specific architectural windows
    \cite{zhao2022surfacecode,takeda2022silicon,acharya2023suppressing,
        google2025belowthreshold}. Such demonstrations make decoder benchmarking
    engineering-relevant, but they also expose a methodological risk: comparisons can
    change when transferred between toolchains or data pipelines without replay
    controls. If two decoders are fed streams that are semantically ``similar'' but
    structurally non-identical (event indexing,
    time stamps, empty-event handling, or parser assumptions), observed differences
    can be partly attributable to interface drift rather than decoder policy.
    Because this drift can be subtle and cumulative, stack-level benchmarking needs
    an explicit, machine-checkable replay contract.

    In parallel with discrete-variable progress, continuous-variable and photonic
    routes based on Gottesman--Kitaev--Preskill (GKP) encodings have advanced from
    theoretical proposals toward experimentally grounded platforms
    \cite{gkp2001,fukui2018highthreshold,tzitrin2021static,madsen2022qca,
        aghaeerad2025aurora,larsen2025integratedgkp,somhorst2026photondistillation}.
    GKP programs add a representation step: analog quadrature shifts are digitized
    into syndrome-like events before or during outer-code processing. This creates a
    hybrid interface problem analogous to the surface-code case, but more sensitive
    to preprocessing conventions. When inner-code information is mapped to
    decoder-facing events, digitization choices can confound decoder-policy
    comparisons unless the replay contract is fixed. This motivates studying
    discrete and GKP-like families under the same downstream decoder IO semantics.

    LiDMaS+ provides the contract layer used here
    \cite{wayo2026lidmas,wayo2026unifiedhardwaretodecoderarchitecturehybrid}.
    Its line-oriented decoder IO format allows independent request generators to
    emit streams that can be replayed through a shared decoder matrix with matched
    configuration fields. Separating generation from replay supports two controlled
    comparisons: multiple decoders can ingest the same stream byte-for-byte, and a
    single decoder can ingest streams from multiple software generators under the
    same post-generation controls. The format also exposes parser-level integrity
    metrics (response ratios, parse failures, decoder-name mismatches, warning
    counts), which distinguish decoder effects from contract violations.

    The focus here is comparability, not absolute threshold competition. We compare
    four syndrome-generation stacks (PennyLane, Qiskit, Cirq, and a LiDMaS+
    reference generator) and replay their request streams through one shared decoder
    set (MWPM, UF, BP). The comparison is performed for two families in one
    reproducible protocol: repeated-round surface-style syndromes and digitized-GKP
    syndromes. The question is narrower than ``which family has the better
    threshold?'': under matched replay semantics, which decoder ordering is stable
    within each family, and how do source-dependent effects change between families?

    We avoid raw cross-family threshold-equivalence claims. Surface and GKP streams
    differ in physical meaning, event-rate profile, syndrome sparsity, and
    preprocessing path. Raw metric magnitudes therefore mix several factors and can
    be misread as direct quality orderings. We instead evaluate decoder separability
    within each family and report cross-family behavior through normalized trends
    that preserve ordering without asserting threshold parity.

    The same issue appears when benchmarking moves from single-lab datasets to
    multi-source data ecosystems. In both
    superconducting and photonic programs, published results are now commonly
    accompanied by heterogeneous artifacts: simulator outputs, processed syndrome
    tables, calibration metadata, and repository snapshots across different tools
    \cite{krinner2022repeated,ni2023breakeven,aghaeerad2025aurora,larsen2025integratedgkp}.
    These practices improve transparency but make comparability fragile when
    downstream analyses merge assets with inconsistent schema assumptions. A replay
    contract treats this as an interface problem: decoder-policy conclusions require
    request and response streams that pass shared integrity bounds.

    The contributions are methodological and empirical. The protocol provides
    family-aware replay with explicit integrity checks, released run artifacts, and
    hidden-truth sidecars for evaluating corrections on a fixed outer-code logical
    path. Empirical outputs include decoder ordering, source-vs-reference deltas,
    confidence intervals, and logical-parity error rates under matched controls for
    two code families. The protocol can be extended to larger shot windows, denser
    parameter grids, learned decoders \cite{gu2026scalableneural}, and
    hardware-adjacent datasets.

    The study targets three goals:
    \begin{enumerate}[leftmargin=*]
  \item verify parser and replay integrity across independent request-generation
  stacks;
  \item quantify decoder-policy separability within each code family and assess
  how source-dependent effects transfer across families;
  \item connect correction volume to an outer-code logical-parity diagnostic without
  making threshold-equivalence claims.
    \end{enumerate}

    \section{Comparative Architecture}
    \label{sec:architecture}
    The execution flow has three stages:
    \begin{enumerate}[leftmargin=*]
  \item generate request streams from each source stack;
  \item replay every request stream through the same decoder set;
  \item aggregate within-family and cross-family metrics under one schema.
    \end{enumerate}

\begin{figure*}[t]
        \centering
        \includegraphics[width=0.6\textwidth]{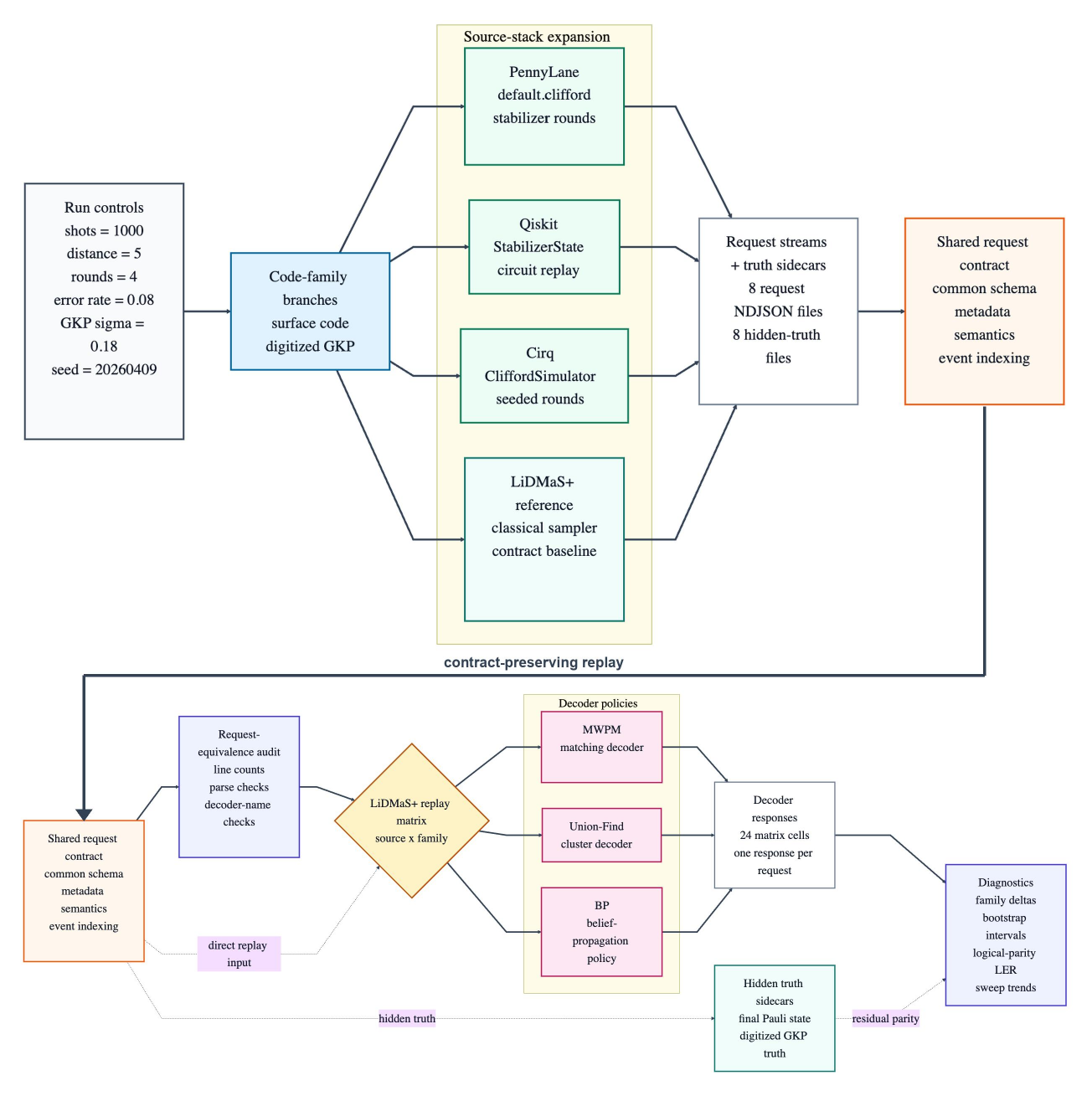}
    \caption{Study architecture. Fixed run controls feed surface-code and digitized-GKP branches; PennyLane, Qiskit, Cirq, and the LiDMaS+ reference then emit request streams and hidden-truth sidecars under one contract. The replay matrix evaluates request streams through the decoder set, while the sidecars are used only after replay for the outer-code logical-parity diagnostic.}
        \label{fig:p4_study_architecture}
    \end{figure*}

    The source stacks are PennyLane, Qiskit, Cirq, and a LiDMaS+ classical
    reference generator. Each source emits request lines into the same decoder IO
    format. Replay then runs MWPM, UF, and BP over every source stream with fixed
    configuration fields outside decoder identity. Figure~\ref{fig:p4_study_architecture}
    summarizes this study architecture from run controls through replay diagnostics.

    \section{Methodology}
    \label{sec:methodology}
    This section defines the replay protocol. The study does not introduce a new
    decoder or code family; it controls interfaces so that decoder-policy differences
    are not mixed with source-specific serialization choices. The protocol proceeds
    from run configuration to request generation, replay execution, family-aware
    aggregation, and uncertainty estimation. Each stage records integrity fields that
    connect request lines to the reported metrics.

    \subsection{Run Configuration}
    The current run uses repeated-round settings with shot count \(=1000\), distance
    \(d=5\), rounds \(=4\), error rate \(=0.08\), and \(\sigma=0.18\). Event export
    uses Z events only (\texttt{emit\_x\_events=0}, \texttt{emit\_z\_events=1}) for
    decoder compatibility across all stacks. Two families are executed:
    \texttt{surface} and \texttt{gkp}. The run seed is \(20260409\), and the replay
    decoder set is \(\{\mathrm{MWPM}, \mathrm{UF}, \mathrm{BP}\}\).
    Source-vs-reference bootstrap analysis uses 5000 response-line resamples per
    metric; rank-stability analysis uses 4000 source-bootstrap draws.

    \subsection{Protocol Structure and Contract Controls}
    The protocol has three boundaries: (i) request construction, (ii) decoder
    replay, and (iii) metric aggregation. At boundary (i), each source stack
    generates line-oriented request records with a fixed field schema and common
    metadata semantics. At boundary (ii), each decoder consumes those records without
    source-specific adapters that alter event meaning. At boundary (iii), all metrics
    are computed from parsed response records under one family-aware table schema.

    We enforce four integrity controls before interpreting policy differences. First,
    line preservation is checked through request/response cardinality and response
    ratio \(\rho\). Second, parser consistency is checked through request/response
    parse-error counts. Third, decoder-identity consistency is checked by matching
    declared decoder labels in responses to the replay target. Fourth, warning-path
    consistency is tracked through warning-no-syndrome counts and rates. The first
    three controls gate strict comparability; the fourth is retained as an
    interpretable behavioral signal rather than an immediate exclusion criterion,
    because warning incidence can still be informative when parser integrity remains
    intact.

    In repeated-round runs, a shot emits a temporal event stream over \(R\) rounds
    rather than a single syndrome snapshot. For each round, source backends compute
    check outcomes; event records are then derived from round-to-round syndrome
    differences. All replay inputs are event-level objects rather than raw state
    vectors. The configuration exports Z events only, which keeps the decoder
    interface common across stacks while retaining repeated-round structure.
    Algorithm~\ref{alg:p4_request_generation} summarizes request construction.

    \begin{algorithm}[t]
\caption{Family-Aware Request Construction Under a Fixed Event Contract}
\label{alg:p4_request_generation}
\KwInput{family \(f\in\{\text{surface},\text{gkp}\}\), source stack \(s\), distance \(d\), rounds \(R\), shots \(N\), noise parameters \(\Theta\), event mask \((m_X,m_Z)\)}
\KwOutput{request stream \(Q_{f,s}=\{q_1,\ldots,q_N\}\)}
Initialize geometry \(G(d)\), pseudo-random state, and serializer fields\;
\ForLoop{\(n \leftarrow 1\) to \(N\)}{
    Initialize latent physical state for family \(f\)\;
    Initialize previous round checks \((\mathbf{s}^{X}_{0},\mathbf{s}^{Z}_{0})\leftarrow \mathbf{0}\)\;
    Initialize empty event list \(E_n\)\;
    \ForLoop{\(r \leftarrow 1\) to \(R\)}{
        Apply family-specific state-noise update using \(\Theta\)\;
        Evaluate round checks \((\hat{\mathbf{s}}^{X}_{r},\hat{\mathbf{s}}^{Z}_{r})\) on source \(s\)\;
        Apply measurement-noise channel to \((\hat{\mathbf{s}}^{X}_{r},\hat{\mathbf{s}}^{Z}_{r})\)\;
        Compute temporal differences \(\Delta \mathbf{s}^{X}_{r}=\hat{\mathbf{s}}^{X}_{r}\oplus \mathbf{s}^{X}_{r-1}\), \(\Delta \mathbf{s}^{Z}_{r}=\hat{\mathbf{s}}^{Z}_{r}\oplus \mathbf{s}^{Z}_{r-1}\)\;
        Emit events selected by \((m_X,m_Z)\) and append to \(E_n\)\;
        Update previous checks to \((\mathbf{s}^{X}_{r},\mathbf{s}^{Z}_{r})\leftarrow(\hat{\mathbf{s}}^{X}_{r},\hat{\mathbf{s}}^{Z}_{r})\)\;
    }
    Serialize one contract-compliant request line \(q_n\) with \(E_n\), noise metadata, and run metadata\;
}
Return \(Q_{f,s}\)\;
    \end{algorithm}

    \subsection{Replay Matrix and Family-Aware Aggregation}
    For each family \(f\), source \(s\), and decoder \(k\), the protocol constructs
    one replay cell \((f,s,k)\). In each cell, the decoder receives the same request
    sequence generated for that \((f,s)\) pair; only decoder identity changes across
    cells. This yields a structured matrix that supports two orthogonal comparisons:
    decoder-separable analysis at fixed source, and source-separable analysis at
    fixed decoder.

    Aggregation is performed after replay parsing, not during replay execution. This
    separates execution failures from metric logic. After parsing, each row exposes
    line-level counts, warning indicators, and correction-set summaries. Family-wise
    tables average over non-reference sources; source-vs-reference tables use paired
    subtraction at fixed \((f,k)\). The pairing avoids mixing decoder effects with
    family or source marginals.

    Uncertainty is estimated with paired bootstrap resampling inside each family.
    For each \((f,s,k)\) versus \((f,\mathrm{ref},k)\) comparison, we resample
    response lines with replacement, recompute mean flip differences, and estimate
    empirical 95\% intervals and one-sided probabilities. We use the bootstrap here
    because closed-form variance approximations for non-Gaussian correction-set
    statistics can be misleading at moderate sample sizes. Bootstrap outputs are
    interval summaries for the current operating point, not decoder constants. In
    addition, request
    generation writes hidden truth sidecars that are never exposed to the decoder.
    These records store the final outer-code X-channel state and a fixed midline
    logical path; after replay, decoder corrections are compared with this hidden
    truth to estimate an outer-code logical-parity error rate.
    Algorithm~\ref{alg:p4_replay_aggregation} summarizes the replay and aggregation
    stage.

    Cross-family interpretation uses normalized trends rather than raw-magnitude
    comparison. Equation~\eqref{eq:p4_normalized_metric} defines this normalization:
    for each family and each metric \(m\), decoder values are mapped to \([0,1]\)
    across decoders in that family,
    \begin{align}
        \tilde{m}_{f,k}=\frac{m_{f,k}-\min_{k'} m_{f,k'}}{\max_{k'} m_{f,k'}-\min_{k'} m_{f,k'}}.
        \label{eq:p4_normalized_metric}
    \end{align}
    This preserves within-family ordering without treating absolute cross-family
    magnitudes as threshold-equivalent.

    \begin{algorithm}[t]
\caption{Contract-Preserving Replay and Family-Aware Metric Aggregation}
\label{alg:p4_replay_aggregation}
\KwInput{families \(\mathcal{F}\), sources \(\mathcal{S}\), decoders \(\mathcal{K}\), reference source \(s_{\mathrm{ref}}\), bootstrap samples \(B\)}
\KwOutput{replay matrix table, family summary table, source-vs-reference table, logical-parity table, normalized cross-family table}
\ForEach{\(f \in \mathcal{F}\)}{
    \ForEach{\(s \in \mathcal{S}\)}{
        \ForEach{\(k \in \mathcal{K}\)}{
            Replay \(Q_{f,s}\) with decoder \(k\) and parse responses\;
            Compute integrity fields and primary metrics \((N^{\mathrm{req}},N^{\mathrm{resp}},\rho,f,\eta,\lambda)\)\;
            Append row for cell \((f,s,k)\)\;
        }
    }
}
Verify matrix-level integrity constraints and record warning incidences\;
\ForEach{\(f \in \mathcal{F}, k \in \mathcal{K}\)}{
    Compute source-mean summaries over \(s\neq s_{\mathrm{ref}}\)\;
    \ForEach{\(s \in \mathcal{S}\setminus\{s_{\mathrm{ref}}\}\)}{
        Compute \(\Delta f_{f,s,k}=f_{f,s,k}-f_{f,s_{\mathrm{ref}},k}\)\;
        Estimate bootstrap interval and tail probability using \(B\) paired resamples\;
    }
}
Normalize selected metrics within each family across decoders\;
Write unified tables and figure-ready summaries\;
    \end{algorithm}

    \subsection{Metrics}
    For dataset/source \(s\), family \(f\), and decoder \(k\), let
    \(N^{\mathrm{req}}_{f,s,k}\) and \(N^{\mathrm{resp}}_{f,s,k}\) be request and
    response line counts. Equations~\eqref{eq:p4_response_ratio},
    \eqref{eq:p4_average_flip_count}, \eqref{eq:p4_nonempty_flip_rate}, and
    \eqref{eq:p4_logical_error_rate} define the integrity, intervention, and
    logical-parity metrics used in the replay tables:
    \begin{align}
        \rho_{f,s,k} &= \frac{N^{\mathrm{resp}}_{f,s,k}}{N^{\mathrm{req}}_{f,s,k}}
        \label{eq:p4_response_ratio}
        && \text{(response ratio)} \\
        f_{f,s,k} &= \frac{1}{N^{\mathrm{resp}}_{f,s,k}}
        \sum_{i=1}^{N^{\mathrm{resp}}_{f,s,k}} \left|\mathrm{flips}_i\right|
        \label{eq:p4_average_flip_count}
        && \text{(average flip count)} \\
        \eta_{f,s,k} &= \frac{N^{\mathrm{nonempty\_flip}}_{f,s,k}}{N^{\mathrm{resp}}_{f,s,k}}
        \label{eq:p4_nonempty_flip_rate}
        && \text{(nonempty flip rate)}.
    \end{align}
    Let \(L_f\) denote the fixed outer-code midline logical path for family \(f\).
    For response \(i\), let \(\mathbf{e}_{f,s,i}\) be the hidden final X-channel
    truth vector and \(\mathbf{c}_{f,s,k,i}\) be the decoder X-correction vector.
    The outer-code logical-parity error rate is
    \begin{align}
        \lambda_{f,s,k} =
        \frac{1}{N^{\mathrm{resp}}_{f,s,k}}
        \sum_{i=1}^{N^{\mathrm{resp}}_{f,s,k}}
        \mathds{1}\!\left[
                        \left((\mathbf{e}_{f,s,i}\oplus\mathbf{c}_{f,s,k,i})\cdot L_f\right)
                        \bmod 2 = 1
        \right].
        \label{eq:p4_logical_error_rate}
    \end{align}
    For the surface family, \(\mathbf{e}\) is the final simulated Pauli X-error
    state. For the GKP family, \(\mathbf{e}\) is the canonical digitized
    shift-derived X-error state. Thus \(\lambda\) is a replay-consistent logical
    parity check for the tested operating point, not a threshold estimate.

    Source-vs-reference effects and family-level source means are reported using
    Eqs.~\eqref{eq:p4_source_delta} and~\eqref{eq:p4_source_mean}:
    \begin{align}
        \Delta f_{f,s,k} = f_{f,s,k} - f_{f,\mathrm{ref},k}.
        \label{eq:p4_source_delta}
    \end{align}
    \begin{align}
        \bar{f}_{f,k}^{\mathrm{src}}=
        \frac{1}{|\mathcal{S}_{\neg \mathrm{ref}}|}
        \sum_{s\in \mathcal{S}_{\neg \mathrm{ref}}} f_{f,s,k},
        \label{eq:p4_source_mean}
    \end{align}
    where \(\mathcal{S}_{\neg \mathrm{ref}}\) excludes the designated reference
    stream. Bootstrap intervals for \(\Delta f_{f,s,k}\) are estimated per family
    with paired resampling under fixed \((f,s,k)\) versus
    \((f,\mathrm{ref},k)\) comparisons.

    \section{Results}
    \label{sec:results}
    \subsection{Replay Integrity}
    The replay contract closed across the full comparison matrix. The unified run
    contains 24 replay cells (4 sources \(\times\) 3 decoders \(\times\) 2 families);
    all 24,000 request lines are matched by 24,000 response lines. The response
    ratio \(\rho\) is \(1.0\) in every cell, so no source-decoder pair is partially
    evaluated or dropped before aggregation.

    Request parse errors, response parse errors, and decoder-name mismatches are all
    zero (Table~\ref{tab:p4_contract}). This matters because parser drift can create
    row-level misalignment that later appears as a decoder effect. Here, the replay
    layer contributes no detected parse or naming failures to the comparison matrix.

    The only nonzero integrity flag is warning-no-syndrome count: 15 total, all in
    GKP-Cirq cells. Relative to the 3,000 lines in that source-family slice, this is
    a low warning rate (\(0.5\%\) per decoder row). Its concentration, not its
    magnitude, is the relevant feature. It marks GKP-Cirq as a source-specific edge
    case rather than a global replay failure.

    \begin{table}[t]
        \centering
        \caption{Contract integrity summary for the unified family-aware replay matrix.}
        \label{tab:p4_contract}
        \begin{tabular}{lr}
            \toprule
            Quantity & Value \\
            \midrule
            Families & 2 \\
            Sources & 4 \\
            Decoders & 3 \\
            Replay cells & 24 \\
            Request lines & 24\,000 \\
            Response lines & 24\,000 \\
            Request parse errors & 0 \\
            Response parse errors & 0 \\
            Decoder mismatches & 0 \\
            Warning-no-syndrome count & 15 \\
            \bottomrule
        \end{tabular}
    \end{table}

    \subsection{Within-Family Decoder Ordering}
\begin{figure*}[t]
        \centering
        \includegraphics[width=0.72\textwidth]{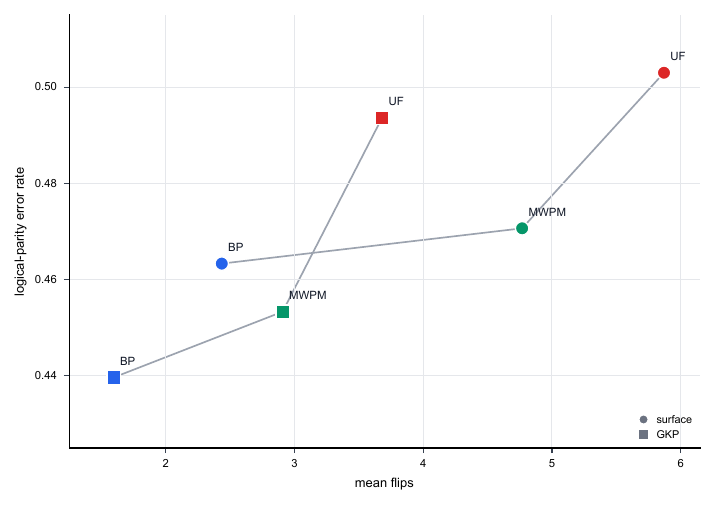}
        \caption{Within-family decoder tradeoff. Each point reports source-mean flip count against the source-mean outer-code logical-parity error rate; circles denote surface rows and squares denote GKP rows. Decoder ordering remains consistent within both code families while operating ranges differ by family.}
        \label{fig:p4_family_tradeoff}
    \end{figure*}

    With replay integrity established, decoder ordering is stable within both code
    families:
    \(\mathrm{BP} < \mathrm{MWPM} < \mathrm{UF}\) by source-averaged flip count
    (Table~\ref{tab:p4_family_decoder_summary}, Figure~\ref{fig:p4_family_tradeoff}).
    For surface data, source-mean values are BP \(2.435\), MWPM \(4.768\), and
    UF \(5.870\). For GKP data, they are BP \(1.595\), MWPM \(2.908\), and
    UF \(3.681\). Thus the ordering is not a single-source artifact; it persists
    after source averaging in both families.

    BP is the lowest-intervention policy in this operating window. Relative to MWPM,
    it reduces mean intervention volume by \(48.9\%\) in surface and \(45.1\%\) in
    GKP; relative to UF, the reductions are \(58.5\%\) and \(56.7\%\). MWPM and UF
    are separated by a smaller but systematic margin (surface: \(4.768 \rightarrow
    5.870\); GKP: \(2.908 \rightarrow 3.681\)).

    The family-level sensitivity column \(\mathrm{mean}\,|\Delta f|\) in
    Table~\ref{tab:p4_family_decoder_summary} adds a second layer. In both
    families, \(\mathrm{mean}\,|\Delta f|\) increases from BP to MWPM to UF
    (surface: \(0.046, 0.051, 0.087\); GKP: \(0.428, 0.682, 0.890\)). In this run,
    higher-intervention decoders also move more when the source generator changes.
    This coupling makes source sensitivity part of decoder selection, alongside
    intervention count and implementation cost.

    \begin{table}[t]
        \centering
        \caption{Family-wise decoder summary for the unified replay matrix.}
        \label{tab:p4_family_decoder_summary}
        \begin{tabular}{lcc}
            \toprule
            Family / Decoder & Source-mean flip & Mean \(|\Delta f|\) vs ref \\
            \midrule
            Surface BP & 2.435 & 0.046 \\
            Surface MWPM & 4.768 & 0.051 \\
            Surface UF & 5.870 & 0.087 \\
            GKP BP & 1.595 & 0.428 \\
            GKP MWPM & 2.908 & 0.682 \\
            GKP UF & 3.681 & 0.890 \\
            \bottomrule
        \end{tabular}
    \end{table}

    \subsection{Source-vs-Reference Effects}
\begin{figure*}[t]
        \centering
        \includegraphics[width=\textwidth]{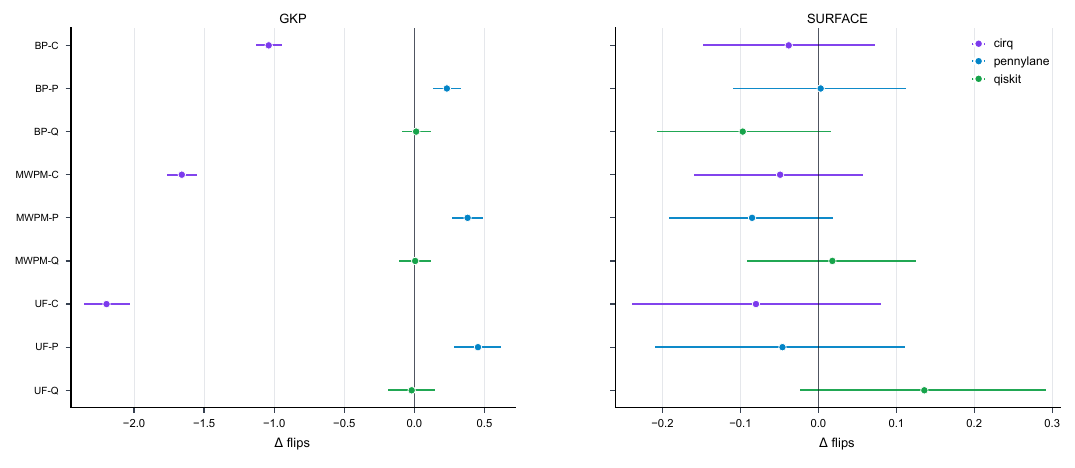}
        \caption{Family-specific source-vs-reference effect sizes with confidence intervals. GKP-Cirq deviations are stronger than surface-Cirq deviations.}
        \label{fig:p4_family_delta_forest}
    \end{figure*}

    Source-vs-reference deltas separate small source offsets from systematic shifts.
    Surface-family deltas are modest: for BP they range from \(-0.097\) (Qiskit) to \(+0.003\)
    (PennyLane), for MWPM from \(-0.085\) (PennyLane) to \(+0.018\) (Qiskit), and
    for UF from \(-0.080\) (Cirq) to \(+0.136\) (Qiskit). Most corresponding
    bootstrap intervals overlap zero, which indicates that surface source effects in
    this run are directionally visible but limited in magnitude
    (Figure~\ref{fig:p4_family_delta_forest}).

    GKP shows a sharper, structured pattern. The Cirq stream is strongly negative
    relative to reference across all decoders:
    \(-1.039\) (BP), \(-1.659\) (MWPM), \(-2.196\) (UF). For all three, the 95\%
    bootstrap intervals remain strictly below zero, so the negative direction is
    stable under resampling. In contrast, PennyLane is strongly positive in GKP:
    \(+0.232\) (BP), \(+0.380\) (MWPM), \(+0.454\) (UF), with 95\% intervals above
    zero for each decoder. Qiskit sits near the reference line in GKP, with small
    deltas (\(+0.013\), \(+0.006\), \(-0.019\)) and intervals crossing zero.

    This three-part GKP structure (Cirq below reference, PennyLane above reference,
    Qiskit near reference) shows that source choice can shift the operating point
    even when decoder and replay semantics are fixed. The shift is coherent across
    decoder families, which points to source-level event statistics rather than
    random decoder instability. The GKP-Cirq warning concentration is consistent with
    this interpretation, although it does not by itself explain the full shift.

\begin{figure*}[t]
        \centering
        \includegraphics[width=\textwidth]{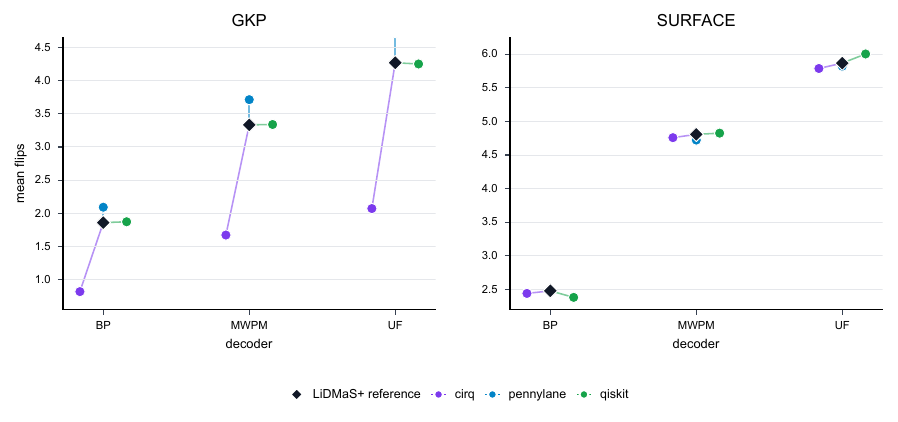}
        \caption{Source-vs-reference decoder comparison in the unified family-aware run. Each panel reports mean flip count by decoder, with the LiDMaS+ reference shown as a diamond and non-reference source stacks shown as colored points linked to the reference operating point.}
        \label{fig:p4_source_vs_lidmas}
    \end{figure*}

    Decoder ranking is stable within a given source, but absolute magnitude can move
    across sources, especially in GKP. Source-vs-reference deltas are therefore
    primary outputs rather than secondary diagnostics.
    Figure~\ref{fig:p4_source_vs_lidmas} preserves
    family-decoder granularity; aggregated totals alone would hide the sign-symmetric
    behavior of Cirq versus PennyLane in GKP.

    The analysis supports a two-step comparison: choose decoder policy using
    within-family ordering under matched semantics, then evaluate source portability
    with \(\Delta f\) and interval structure before transferring conclusions across
    generation stacks.

    \subsection{Cross-Family Normalized Trends}
\begin{figure*}[t]
        \centering
        \includegraphics[width=0.78\textwidth]{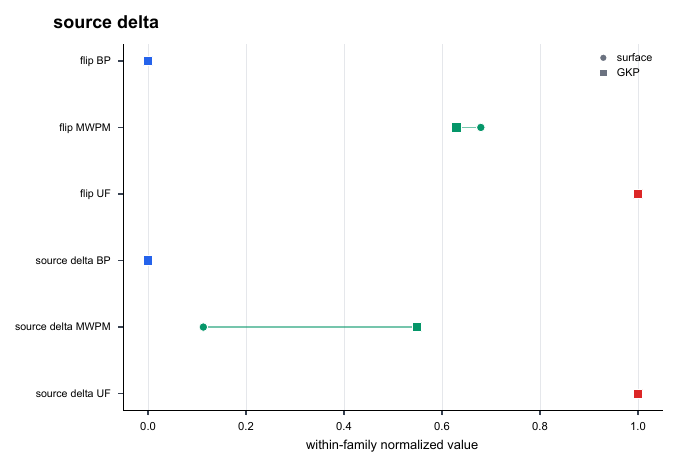}
        \caption{Cross-family normalized trends. Ordering is compared across families using normalized metrics rather than raw cross-family threshold equivalence.}
        \label{fig:p4_cross_family_normalized}
    \end{figure*}

    Raw magnitudes are not compared across families as if they represented a
    single threshold axis. The normalized trend view
    (Figure~\ref{fig:p4_cross_family_normalized}) addresses this by comparing decoder
    position within each family rather than asserting direct surface-to-GKP
    equivalence. On normalized flip magnitude, both families have the same monotone
    trajectory: BP at \(0.0\), MWPM at intermediate levels (surface \(0.679\), GKP
    \(0.629\)), and UF at \(1.0\).

    The normalized stack-delta metric shows a second-order difference between
    families: MWPM occupies a much larger fraction of the BP-to-UF spread in GKP
    (\(0.549\)) than in surface (\(0.113\)). In GKP, MWPM is therefore a distinct
    point on the intervention/sensitivity axis; in surface, it is closer to BP in
    source-delta positioning.

    The normalized warning component is flat (\(0.5\) for all rows) because warning
    rates are near-zero and nearly identical in this run, aside from the localized
    GKP-Cirq slice. Warnings are therefore integrity checks and anomaly markers, not
    the main separator of decoder behavior. The separating signals are average flip
    magnitude and source-vs-reference displacement.

    \subsection{Logical-Parity Error Rate}
    Hidden-truth sidecars add a second outcome layer. After replay, each decoder
    correction is combined with the final generated X-channel truth and tested on a
    fixed outer-code midline logical path. This gives the logical-parity error rate
    \(\lambda\) in Eq.~\eqref{eq:p4_logical_error_rate}. The metric is narrower than
    a threshold-style logical error-rate study, but it links correction volume to a
    residual logical-parity outcome under the same replay contract.

\begin{figure*}[t]
        \centering
        \includegraphics[width=0.72\textwidth]{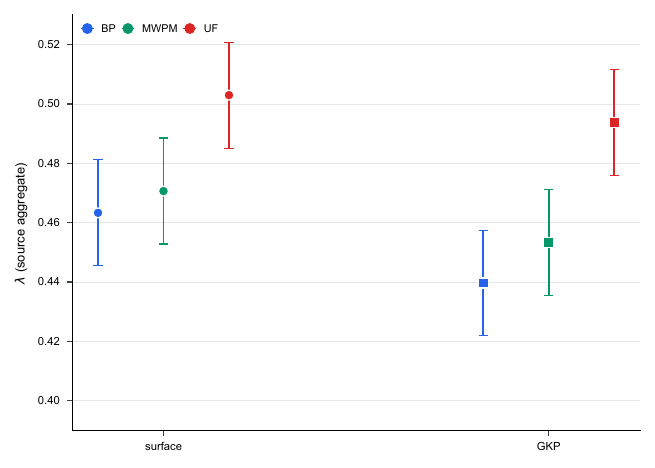}
        \caption{Outer-code logical-parity error rate from hidden truth sidecars. Points report source-aggregated \(\lambda\) for each family and decoder; intervals are Wilson 95\% binomial intervals. The metric evaluates residual parity on a fixed midline logical path and is used as a comparative replay diagnostic, not as a threshold estimate.}
        \label{fig:p4_logical_error_rate}
    \end{figure*}

    Figure~\ref{fig:p4_logical_error_rate} shows that source-mean \(\lambda\)
    broadly follows the same decoder ordering as intervention volume. In surface,
    source-mean rates are BP \(0.463\), MWPM \(0.471\), and UF \(0.503\). In GKP,
    they are BP \(0.440\), MWPM \(0.453\), and UF \(0.494\). The BP--MWPM
    separation is modest, while UF is consistently the largest source-mean rate in
    both families. In this operating point, lower intervention is not accompanied by
    a higher logical-parity residual.

    The absolute magnitudes of \(\lambda\) are limited by the run design: one
    high-noise operating point, one fixed logical path, and one repeated-round event
    contract. The rates are comparative replay diagnostics; they do not replace a
    logical-threshold sweep over distance, rounds, and noise.

    \subsection{Rank Stability and Variance Attribution}
\begin{figure*}[t]
        \centering
        \includegraphics[width=0.72\textwidth]{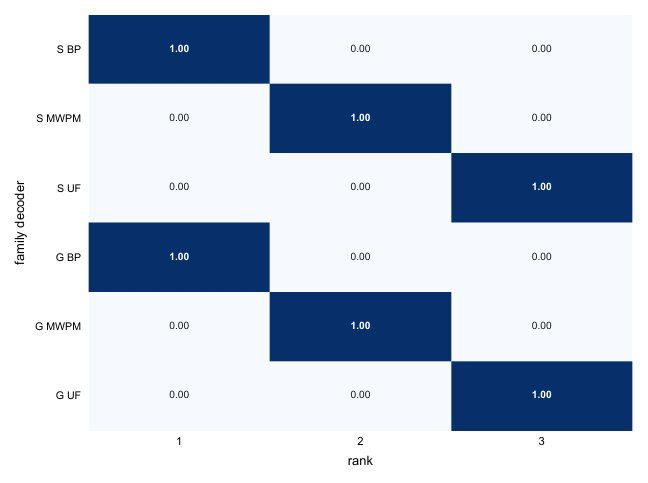}
        \caption{Source-bootstrap decoder rank stability within each code family. Each cell reports the probability that a decoder occupies a rank after resampling the non-reference source stacks and recomputing source-averaged flip count.}
        \label{fig:p4_rank_stability}
    \end{figure*}

    We tested decoder ordering with source-bootstrap rank analysis within each
    family. Each bootstrap
    draw resamples the three non-reference source stacks and recomputes the decoder
    order by source-averaged flip count. The rank pattern is stable in this run:
    BP has rank-1 probability \(1.00\), MWPM has rank-2 probability \(1.00\), and
    UF has rank-3 probability \(1.00\) in both surface and GKP
    (Figure~\ref{fig:p4_rank_stability}). This does not imply rank invariance outside
    the tested operating window, but it shows that the observed ordering is not
    sensitive to which non-reference generator is sampled in the current matrix.

\begin{figure*}[t]
        \centering
        \includegraphics[width=0.82\textwidth]{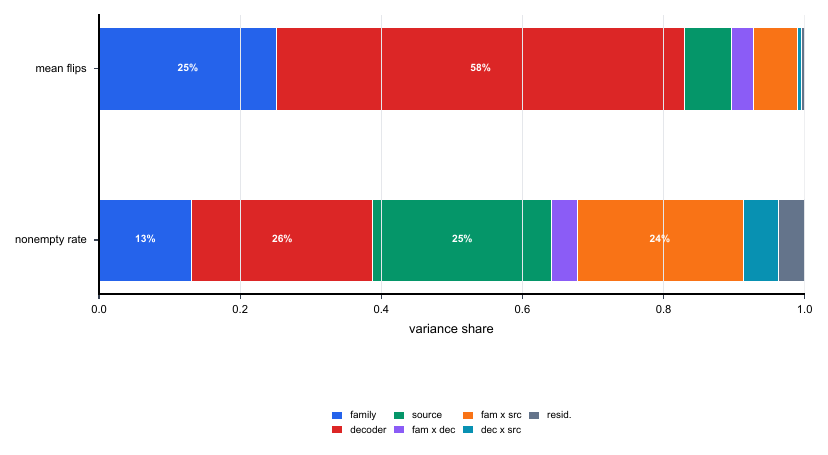}
        \caption{Balanced variance attribution across the unified replay matrix. Shares are computed from the \(2\)-family, \(4\)-source, \(3\)-decoder aggregate matrix and report how much variation in each metric is associated with family, decoder, source stack, and interaction terms.}
        \label{fig:p4_variance_decomposition}
    \end{figure*}

    We also decomposed the balanced \(2 \times 4 \times 3\) matrix into family,
    decoder, source-stack, and interaction components
    (Figure~\ref{fig:p4_variance_decomposition}). For average flip count, decoder
    identity explains the largest share of variation (\(57.8\%\)), followed by
    family (\(25.2\%\)); source-stack main effects account for \(6.7\%\), and the
    family-by-source interaction accounts for \(6.1\%\). Decoder policy therefore
    dominates intervention volume, while family and source conditioning remain
    secondary.

    The nonempty-flip-rate decomposition is less decoder dominated. Decoder
    accounts for \(25.7\%\), source stack for \(25.4\%\), and family-by-source
    interaction for \(23.5\%\), with family itself contributing \(13.1\%\). This is
    consistent with the GKP source-effect result: source identity and
    family-dependent source behavior matter more for event occupancy than for average
    correction volume. We treat this decomposition descriptively rather than
    causally, because each matrix cell is one aggregate replay condition.

    \subsection{Scaling and Parametric Checks}
    Sweeps are separate from the 1000-shot unified
    matrix and are interpreted only as consistency probes, not as replacements for
    the main matrix estimates. Across tested points, they preserve the same decoder
    ordering found in the core results. In the distance-sweep sample
    (\(d=3\), shots \(=30\)), average flip counts are BP \(0.75\), MWPM
    \(1.35\), UF \(1.47\). In the noise-round sample (\(p=0.08\), rounds \(=2\),
    shots \(=30\)), the ordering remains BP \(2.31\), MWPM \(3.96\), UF \(5.13\).

    These checks show that the ranking is not tied to one exact parameter point. They
    also motivate keeping the 1000-shot unified matrix as the quantitative anchor:
    small-sample probes establish directionality, but cross-source interpretation
    requires the larger, contract-verified replay grid.

    \section{Discussion}
    \label{sec:discussion}
    Comparability is the condition for interpreting the decoder results. The replay
    matrix achieved exact line-level closure (24,000 request lines and 24,000
    response lines), with zero parse failures and zero decoder-name mismatches. This
    does not prove decoder quality, but it removes a common source of ambiguity in
    multi-stack benchmarking. Schema and parser differences are not left as hidden
    confounders; they are reported as diagnostics.

    Under this contract, the main behavioral result is the within-family ordering
    \( \mathrm{BP} < \mathrm{MWPM} < \mathrm{UF} \) for average flip count. The
    ordering appears in surface and GKP under the same replay semantics. For the
    tested regime, BP is the lowest-intervention option, MWPM is intermediate, and
    UF is highest.

    Source-bootstrap resampling leaves the decoder ranking unchanged in both
    families, so the ordering claim does not depend on one generator slice. The
    variance decomposition shows that average correction volume is primarily
    decoder-driven, whereas nonempty correction occupancy depends more on decoder,
    source stack, and family-by-source interaction. The logical-parity check gives a
    residual-outcome comparison: under the fixed midline path, UF also has the
    largest source-mean residual parity rate in both families, while BP and MWPM
    remain closer.

    Ranking alone does not capture portability across generation stacks.
    Source-vs-reference effects are family dependent. In surface, source deltas are
    small and mostly centered near zero. In GKP, source effects are larger and
    directional: Cirq is strongly negative relative to reference for all decoders,
    PennyLane is strongly positive, and Qiskit is near reference. The GKP comparison
    is therefore not a rescaled copy of the surface comparison; the digitized path
    introduces stronger source shaping, likely through the mapping from inner-code
    information to outer decoder-facing events.

    Table~\ref{tab:p4_discussion_comparison} summarizes this contrast. Mean
    \(|\Delta f|\) is larger in GKP than in surface for every decoder, so source
    movement is family-scaled rather than decoder-specific. CI-stable source shifts
    are absent in surface and present in GKP (2 out of 3 non-reference sources for
    each decoder), showing that the GKP sign pattern is stable under bootstrap
    resampling.

    \begin{table*}[t]
        \centering
        \caption{Family-conditioned decoder comparison. ``CI-stable source shifts'' counts non-reference sources (out of 3) whose 95\% bootstrap interval for \(\Delta f\) excludes zero.}
        \label{tab:p4_discussion_comparison}
        \begin{tabular}{lcccccc}
            \toprule
            Decoder & Surface \(f\) & GKP \(f\) & Surface \(|\Delta f|\) & GKP \(|\Delta f|\) & Stable shifts, surface & Stable shifts, GKP \\
            \midrule
            BP & 2.435 & 1.595 & 0.046 & 0.428 & 0/3 & 2/3 \\
            MWPM & 4.768 & 2.908 & 0.051 & 0.682 & 0/3 & 2/3 \\
            UF & 5.870 & 3.681 & 0.087 & 0.890 & 0/3 & 2/3 \\
            \bottomrule
        \end{tabular}
    \end{table*}

    The results separate decoder selection into family-conditioned and
    source-conditioned components. In this dataset, BP is the
    lowest-intervention decoder in both families, but source identity controls how
    much margin is needed when moving between generator stacks. In GKP, source
    identity can move mean intervention by more than one flip per response.

    Intervention level and source sensitivity are coupled in this run:
    mean \(|\Delta f|\) grows from BP to MWPM to UF in both families. Selecting a
    decoder therefore also selects a sensitivity profile with respect to source
    variability.

    Normalized cross-family trends preserve within-family ordering without asserting
    that surface and GKP raw magnitudes share a threshold scale. Lower absolute flip
    count in one family does not imply better correction performance in another,
    because event formation and semantics differ. The normalized representation is an
    ordering comparison, not a universal performance metric.

    The pattern is consistent with broader decoder literature in one specific
    respect: decoder tradeoffs are regime contingent, and conclusions are strongest
    when tied to explicit data-generation assumptions
    \cite{duclos2010fast,unionfind2017,pymatching2021,higgott2023improved,
        skoric2023parallel,tan2023scalable}.
    For GKP-oriented settings, the data are consistent with a known architectural
    principle: inner-to-outer translation choices affect the downstream workload seen
    by discrete decoders
    \cite{gkp2001,fukui2018highthreshold,tzitrin2021static,madsen2022qca}. This
    analysis does not isolate which preprocessing component causes each shift, but the
    shifts are measurable, directional, and large enough to affect comparisons.

    Several limits bound the interpretation. First, the core matrix is a single
    operating window (fixed distance, rounds, noise, and shot count). While
    optional sweeps preserve ordering at tested points, they do not map the full
    phase space. Second, the study uses one intervention-centric outcome
    (\(f\), average flip count) as the main comparator. That metric captures
    decoder actuation pressure and relative policy behavior; the added
    logical-parity rate links those corrections to one residual logical path, but it
    is not a full threshold estimate. Third, the analysis compares source streams
    after they are produced; it does not decompose causality inside each generator
    stack. They restrict what can be inferred about absolute fault-tolerance
    performance.

    Natural extensions are denser operating grids (distance, rounds, and noise),
    larger shot windows, additional decoder families, and targeted GKP attribution.
    In the last case, specific digitization or preprocessing knobs would be perturbed
    one at a time while replay and decoder configuration remain fixed.

    The contribution is not a universal decoder-winner claim. It is a comparison
    protocol that separates contract integrity, within-family decoder ranking, and
    source portability into measurable layers. Under this protocol, decoder ordering
    is stable across surface and GKP for the tested regime, but source-conditioned
    effect magnitude is larger in GKP than in surface.

    \section{Conclusion}
    \label{sec:conclusion}
    We report a contract-based protocol for decoder comparison across
    heterogeneous quantum software stacks. Across PennyLane, Qiskit, Cirq, and the
    LiDMaS+ reference stream, the unified matrix closed exactly
    (24\,000 request lines, 24\,000 response lines), with response ratio \(=1.0\)
    in every cell, zero request/response parse failures, and zero decoder-name
    mismatches. These integrity checks make the downstream policy comparisons
    interpretable.

    Within that controlled setting, decoder ordering was stable in both families:
    BP \(<\) MWPM \(<\) UF for source-averaged flip count. Surface means were
    2.435, 4.768, and 5.870 (BP, MWPM, UF), while GKP means were 1.595, 2.908, and
    3.681. Relative to MWPM, BP reduced intervention volume by \(48.9\%\) in surface
    and \(45.1\%\) in GKP. For the tested regime, BP minimizes intervention pressure,
    MWPM occupies a middle operating point, and UF is the highest-intervention
    option. Hidden-truth sidecars showed that UF has the largest source-mean
    outer-code logical-parity error rate in both families, while BP and MWPM remain
    closer on that residual-parity diagnostic.

    At the same time, source portability differed by family. Surface
    source-vs-reference deltas were generally small and often centered near zero,
    whereas GKP showed coherent directional shifts across decoders, including strong
    negative Cirq offsets and positive PennyLane offsets relative to reference.
    These effects, together with concentrated warning-no-syndrome events in
    GKP-Cirq rows, show larger cross-source variation in GKP-oriented settings even
    when replay semantics are fixed.

    The main contribution is a contract-preserving, family-aware comparison
    framework that yields stable within-family decoder ranking while quantifying
    source-conditioned movement. It avoids raw cross-family magnitude comparisons and unsupported threshold-equivalence claims. Follow-on studies can increase shot
    budgets, expand operating grids (distance, rounds, noise), add decoder families,
    connect to architecture-level surface-code layouts \cite{ha2025architectures},
    and isolate GKP preprocessing factors through targeted ablations under the same
    replay contract.

    \section*{Author Contributions}
D.D.K.W.: conceptualization, methodology, software, validation, analysis, writing. C.O.: methodology, software, validation, writing. R.A.D.: methodology, software, validation, writing. L.G.: methodology, software, validation, writing.  Z.M.Y: methodology, validation, writing.  S.G.: methodology, software, validation, supervision, writing.

    \section*{Acknowledgments}
    The authors acknowledge contributors and users who provided feedback on decoder replay and comparative analysis tooling.

    \section*{Data \& Code Availability}
    All data products used in this study are contained in
    \texttt{examples/paper\_runs/paper\_04/results/}. Code is available at
    \href{https://github.com/DennisWayo/lidmas_cpp}{https://github.com/DennisWayo/lidmas\_cpp}.

    \section*{Funding}
    No external funding was received.

    \section*{Disclosure statement}
    The authors report no potential conflicts of interest.

    \appendix
    \section{Reproducibility Commands}
    \label{app:p4_repro}
    Run all commands from the repository root. The wrapper
    \texttt{run\_all.sh} regenerates the family runs, hidden-truth sidecars, replay
    matrix, extended analysis, and unified tables/figures. Circuit-atlas and
    indexed-layout figures are generated separately after
    \texttt{results/03\_analysis/runs/} exists.

    \textbf{(Unified results)}
    \begingroup\footnotesize
    \begin{verbatim}
LIDMAS_P4_SHOTS=1000 \
LIDMAS_P4_DISTANCE=5 \
LIDMAS_P4_ROUNDS=4 \
LIDMAS_P4_ERROR_RATE=0.08 \
LIDMAS_P4_SIGMA=0.18 \
LIDMAS_P4_SEED=20260409 \
LIDMAS_P4_BOOTSTRAP=5000 \
LIDMAS_DECODERS=mwpm,uf,bp \
LIDMAS_P4_CODE_FAMILIES=surface,gkp \
./examples/paper_runs/run_all.sh
    \end{verbatim}
    \endgroup

    \textbf{Full run driver}
    \begingroup\footnotesize
    \begin{verbatim}
./examples/paper_runs/run_all.sh
    \end{verbatim}
    \endgroup

    \textbf{Strict three-stack run}
    \begingroup\footnotesize
    \begin{verbatim}
LIDMAS_P4_PENNYLANE_MODE=required \
LIDMAS_P4_QISKIT_MODE=required \
LIDMAS_P4_CIRQ_MODE=required \
./examples/paper_runs/run_all.sh
    \end{verbatim}
    \endgroup

    \textbf{Parametric sweeps}
    \begingroup\footnotesize
    \begin{verbatim}
LIDMAS_P4_ENABLE_PARAM_SWEEPS=1 \
./examples/paper_runs/run_all.sh
    \end{verbatim}
    \endgroup

    \textbf{Appendix circuit-atlas and indexed-layout figures}
    \begingroup\footnotesize
    \begin{verbatim}
P4=examples/paper_runs/paper_04
RUNS=$P4/results/03_analysis/runs
PRINTS=$P4/results/03_analysis/circuit_prints
FIGS=$P4/results/03_analysis/manuscript_figures

mkdir -p .cache/home/.cache .cache/matplotlib
export HOME=.cache/home
export XDG_CACHE_HOME=.cache/home/.cache
export MPLCONFIGDIR=.cache/matplotlib
export MPLBACKEND=Agg

./.venv/bin/python \
$P4/scripts/print_run_circuits.py \
  --run-root $RUNS \
  --out-dir $PRINTS

./.venv/bin/python \
$P4/scripts/render_circuit_atlases.py \
  --print-dir $PRINTS \
  --out-dir $FIGS

./.venv/bin/python \
$P4/scripts/render_2d_syndrome_layouts.py \
  --run-root $RUNS \
  --out-dir $FIGS
    \end{verbatim}
    \endgroup

    \section{Circuit Print Summaries and Excerpts}
    \label{app:p4_circuit_prints}
    This appendix summarizes the circuit/logic print snapshots stored in
    \texttt{results/03\_analysis/circuit\_prints/}. The grouped circuit atlas is
    rendered into \texttt{results/03\_analysis/manuscript\_figures/}; panel counts
    come from the archived page tiling in
    \texttt{results/03\_analysis/circuit\_figures/}. The counts compare the textual
    footprint of each source representation for the same
    \((d=5,\text{ rounds}=4,\text{ shots}=1000)\) run
    (Table~\ref{tab:p4_circuit_print_summary}).

    \begin{table*}[t]
        \centering
        \caption{Circuit-print artifact summary for the current manuscript run. ``Figure panels'' counts archived PNG page panels associated with each print snapshot.}
        \label{tab:p4_circuit_print_summary}
        \begin{tabular}{lccr}
            \toprule
            Artifact stem & Family & Text lines & Figure panels \\
            \midrule
            surface\_pennylane\_circuit & surface & 490 & 3 \\
            surface\_qiskit\_circuit & surface & 162 & 2 \\
            surface\_cirq\_circuit & surface & 162 & 1 \\
            surface\_lidmas\_reference\_logic & surface & 49 & 1 \\
            surface\_check\_supports & surface & 47 & 1 \\
            surface\_run\_metadata & surface & 16 & 1 \\
            gkp\_pennylane\_digitized\_logic & gkp & 59 & 1 \\
            gkp\_qiskit\_digitized\_logic & gkp & 59 & 1 \\
            gkp\_cirq\_digitized\_logic & gkp & 59 & 1 \\
            gkp\_lidmas\_reference\_digitized\_logic & gkp & 59 & 1 \\
            gkp\_check\_supports & gkp & 47 & 1 \\
            gkp\_run\_metadata & gkp & 16 & 1 \\
            \bottomrule
        \end{tabular}
    \end{table*}

\begin{figure*}[p]
        \centering
        \includegraphics[width=\textwidth]{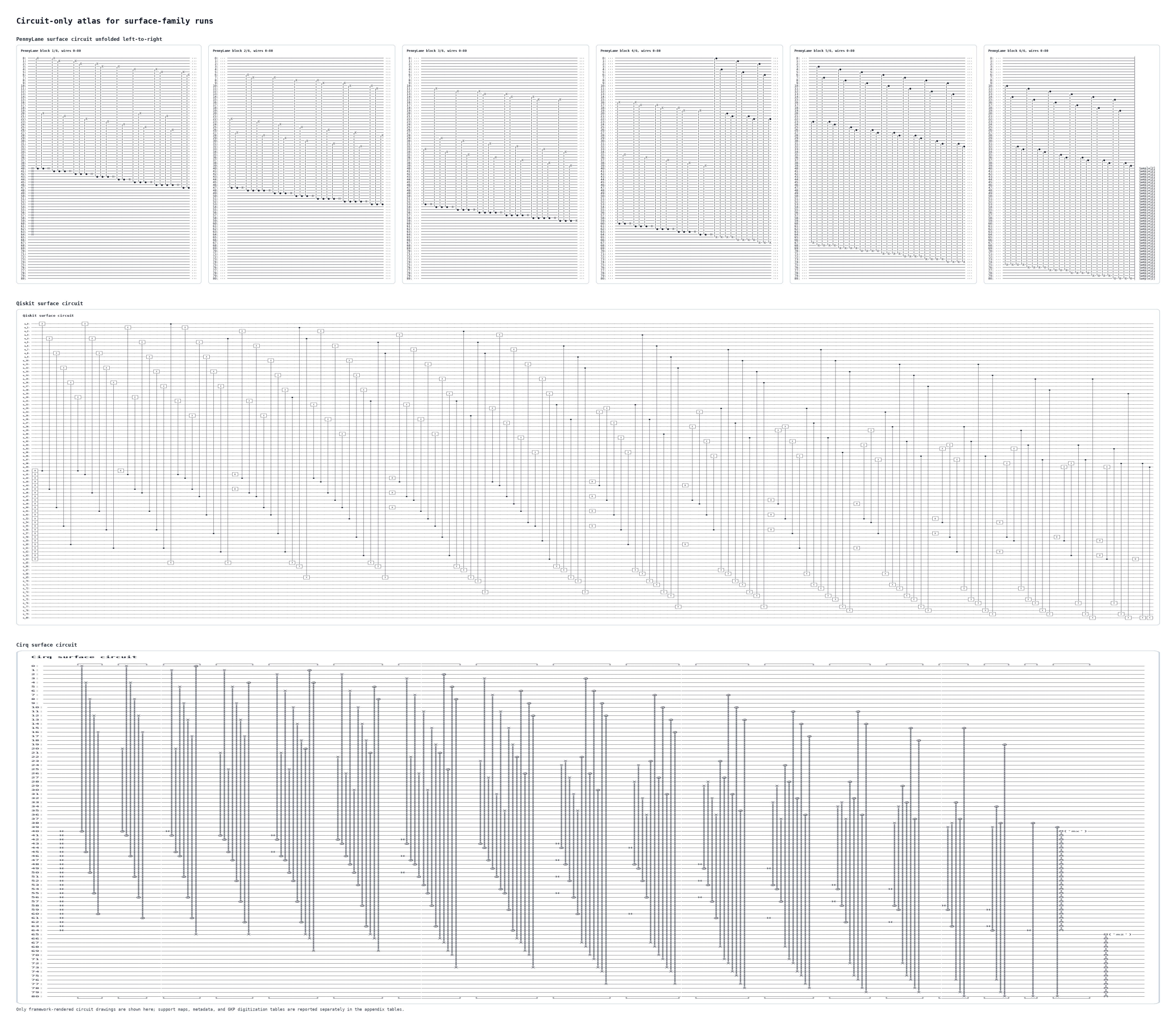}
        \caption{Circuit-only atlas of the surface-family artifacts used in the manuscript run. The PennyLane drawing is unfolded into adjacent wire-\(0\)--\(80\) blocks so that its vertically wrapped text output reads left-to-right; the Qiskit and Cirq circuit drawings are shown as widened single panels.}
        \label{fig:p4_surface_circuit_atlas}
    \end{figure*}

    Here, \texttt{*\_circuit} artifacts are framework text-drawn gate-level
    surface circuits; \texttt{*\_digitized\_logic} artifacts are GKP
    digitization-rule and support snapshots; \texttt{*\_check\_supports} and
    \texttt{*\_run\_metadata} are structural and run-configuration summaries.
    Figure~\ref{fig:p4_surface_circuit_atlas} shows the surface-family circuit
    drawings. GKP digitization rules, support maps, and metadata are reported in the
    tables below.

    The run metadata prints are identical across families except for the family tag
    (\texttt{code\_family = surface} vs. \texttt{code\_family = gkp}): shots
    \(=1000\), rounds \(=4\), distance \(=5\), \(n_{\mathrm{qubits}}=40\),
    \(n_X=25\), \(n_Z=16\), error\_rate \(=0.08\), sigma \(=0.18\), and seed
    \(=20260409\).

    \textbf{Shared lattice-index contract}
    All four source stacks use the same outer-code index convention. The surface
    family realizes it as gate-level check extraction where available; the GKP
    family uses the same outer indices after inner-code digitization. The index
    tables are therefore not repeated per stack.
    Table~\ref{tab:p4_stack_index_contract} summarizes stack usage.

    Tables~\ref{tab:p4_data_index_map}, \ref{tab:p4_check_index_locations}, and
    \ref{tab:p4_check_support_map} give the concrete data-qubit locations,
    check-node locations, and support lists for the \(d=5\) outer graph used in
    both families.

    \begin{table*}[t]
        \centering
        \caption{How the shared lattice-index contract is used by each source stack.}
        \label{tab:p4_stack_index_contract}
        \begin{tabular}{lll}
            \toprule
            Source stack & Surface-family interpretation & GKP-family interpretation \\
            \midrule
            PennyLane & Gate-level surface circuit print & Digitized GKP logic variant using shared outer indices \\
            Qiskit & Gate-level surface circuit print & Digitized GKP logic variant using shared outer indices \\
            Cirq & Gate-level surface circuit print & Digitized GKP logic variant using shared outer indices \\
            LiDMaS+ reference & Classical parity-check reference sampler & Periodic-threshold GKP reference using shared outer indices \\
            \bottomrule
        \end{tabular}
    \end{table*}

    \begin{table*}[t]
        \centering
        \scriptsize
        \caption{Data-qubit index locations for the \(d=5\) outer support graph. Horizontal data qubits occupy \((x+0.5,y)\); vertical data qubits occupy \((x,y+0.5)\).}
        \label{tab:p4_data_index_map}
        \begin{tabular}{lcc lcc}
            \toprule
            Data index & Coordinate & Orientation & Data index & Coordinate & Orientation \\
            \midrule
            D00 & (0.5, 0.0) & horizontal & D20 & (0.0, 0.5) & vertical \\
            D01 & (1.5, 0.0) & horizontal & D21 & (1.0, 0.5) & vertical \\
            D02 & (2.5, 0.0) & horizontal & D22 & (2.0, 0.5) & vertical \\
            D03 & (3.5, 0.0) & horizontal & D23 & (3.0, 0.5) & vertical \\
            D04 & (0.5, 1.0) & horizontal & D24 & (4.0, 0.5) & vertical \\
            D05 & (1.5, 1.0) & horizontal & D25 & (0.0, 1.5) & vertical \\
            D06 & (2.5, 1.0) & horizontal & D26 & (1.0, 1.5) & vertical \\
            D07 & (3.5, 1.0) & horizontal & D27 & (2.0, 1.5) & vertical \\
            D08 & (0.5, 2.0) & horizontal & D28 & (3.0, 1.5) & vertical \\
            D09 & (1.5, 2.0) & horizontal & D29 & (4.0, 1.5) & vertical \\
            D10 & (2.5, 2.0) & horizontal & D30 & (0.0, 2.5) & vertical \\
            D11 & (3.5, 2.0) & horizontal & D31 & (1.0, 2.5) & vertical \\
            D12 & (0.5, 3.0) & horizontal & D32 & (2.0, 2.5) & vertical \\
            D13 & (1.5, 3.0) & horizontal & D33 & (3.0, 2.5) & vertical \\
            D14 & (2.5, 3.0) & horizontal & D34 & (4.0, 2.5) & vertical \\
            D15 & (3.5, 3.0) & horizontal & D35 & (0.0, 3.5) & vertical \\
            D16 & (0.5, 4.0) & horizontal & D36 & (1.0, 3.5) & vertical \\
            D17 & (1.5, 4.0) & horizontal & D37 & (2.0, 3.5) & vertical \\
            D18 & (2.5, 4.0) & horizontal & D38 & (3.0, 3.5) & vertical \\
            D19 & (3.5, 4.0) & horizontal & D39 & (4.0, 3.5) & vertical \\
            \bottomrule
        \end{tabular}
    \end{table*}

    \begin{table}[t]
        \centering
        \scriptsize
        \caption{Check-node index locations for the same \(d=5\) outer support graph. X checks sit on integer lattice vertices; Z checks sit on plaquette centers.}
        \label{tab:p4_check_index_locations}
        \begin{tabular}{lc lc}
            \toprule
            X-check index & Coordinate & Z-check index & Coordinate \\
            \midrule
            X00 & (0.0, 0.0) & Z00 & (0.5, 0.5) \\
            X01 & (1.0, 0.0) & Z01 & (1.5, 0.5) \\
            X02 & (2.0, 0.0) & Z02 & (2.5, 0.5) \\
            X03 & (3.0, 0.0) & Z03 & (3.5, 0.5) \\
            X04 & (4.0, 0.0) & Z04 & (0.5, 1.5) \\
            X05 & (0.0, 1.0) & Z05 & (1.5, 1.5) \\
            X06 & (1.0, 1.0) & Z06 & (2.5, 1.5) \\
            X07 & (2.0, 1.0) & Z07 & (3.5, 1.5) \\
            X08 & (3.0, 1.0) & Z08 & (0.5, 2.5) \\
            X09 & (4.0, 1.0) & Z09 & (1.5, 2.5) \\
            X10 & (0.0, 2.0) & Z10 & (2.5, 2.5) \\
            X11 & (1.0, 2.0) & Z11 & (3.5, 2.5) \\
            X12 & (2.0, 2.0) & Z12 & (0.5, 3.5) \\
            X13 & (3.0, 2.0) & Z13 & (1.5, 3.5) \\
            X14 & (4.0, 2.0) & Z14 & (2.5, 3.5) \\
            X15 & (0.0, 3.0) & Z15 & (3.5, 3.5) \\
            X16 & (1.0, 3.0) &  &  \\
            X17 & (2.0, 3.0) &  &  \\
            X18 & (3.0, 3.0) &  &  \\
            X19 & (4.0, 3.0) &  &  \\
            X20 & (0.0, 4.0) &  &  \\
            X21 & (1.0, 4.0) &  &  \\
            X22 & (2.0, 4.0) &  &  \\
            X23 & (3.0, 4.0) &  &  \\
            X24 & (4.0, 4.0) &  &  \\
            \bottomrule
        \end{tabular}
    \end{table}

    \begin{table}[t]
        \centering
        \scriptsize
        \caption{Complete check-index support map used by the surface and digitized-GKP outer support graph.}
        \label{tab:p4_check_support_map}
        \begin{tabular}{ll ll}
            \toprule
            X check & Data-qubit support & Z check & Data-qubit support \\
            \midrule
            X00 & [0, 20] & Z00 & [0, 4, 20, 21] \\
            X01 & [0, 1, 21] & Z01 & [1, 5, 21, 22] \\
            X02 & [1, 2, 22] & Z02 & [2, 6, 22, 23] \\
            X03 & [2, 3, 23] & Z03 & [3, 7, 23, 24] \\
            X04 & [3, 24] & Z04 & [4, 8, 25, 26] \\
            X05 & [4, 20, 25] & Z05 & [5, 9, 26, 27] \\
            X06 & [4, 5, 21, 26] & Z06 & [6, 10, 27, 28] \\
            X07 & [5, 6, 22, 27] & Z07 & [7, 11, 28, 29] \\
            X08 & [6, 7, 23, 28] & Z08 & [8, 12, 30, 31] \\
            X09 & [7, 24, 29] & Z09 & [9, 13, 31, 32] \\
            X10 & [8, 25, 30] & Z10 & [10, 14, 32, 33] \\
            X11 & [8, 9, 26, 31] & Z11 & [11, 15, 33, 34] \\
            X12 & [9, 10, 27, 32] & Z12 & [12, 16, 35, 36] \\
            X13 & [10, 11, 28, 33] & Z13 & [13, 17, 36, 37] \\
            X14 & [11, 29, 34] & Z14 & [14, 18, 37, 38] \\
            X15 & [12, 30, 35] & Z15 & [15, 19, 38, 39] \\
            X16 & [12, 13, 31, 36] &  &  \\
            X17 & [13, 14, 32, 37] &  &  \\
            X18 & [14, 15, 33, 38] &  &  \\
            X19 & [15, 34, 39] &  &  \\
            X20 & [16, 35] &  &  \\
            X21 & [16, 17, 36] &  &  \\
            X22 & [17, 18, 37] &  &  \\
            X23 & [18, 19, 38] &  &  \\
            X24 & [19, 39] &  &  \\
            \bottomrule
        \end{tabular}
    \end{table}

    \textbf{GKP digitization-rule excerpt (from \texttt{gkp\_pennylane\_digitized\_logic.txt})}
    \begingroup\footnotesize
    \begin{verbatim}
Digitization rules:
  - pennylane: periodic threshold with small bias
  - qiskit:
      rounded scaled value with
      Gaussian perturbation
  - cirq: sinusoidal phase-sign rule
  - lidmas_reference:
      periodic threshold without framework bias
    \end{verbatim}
    \endgroup

    \section{Syndrome Layout and GKP Inner-Code Figures}
    \label{app:p4_layouts}
\begin{figure*}[t]
        \centering
        \subfloat[Surface-code outer support topology with indexed data and check nodes (X-only, Z-only, and combined view).]{
            \includegraphics[width=0.95\textwidth]{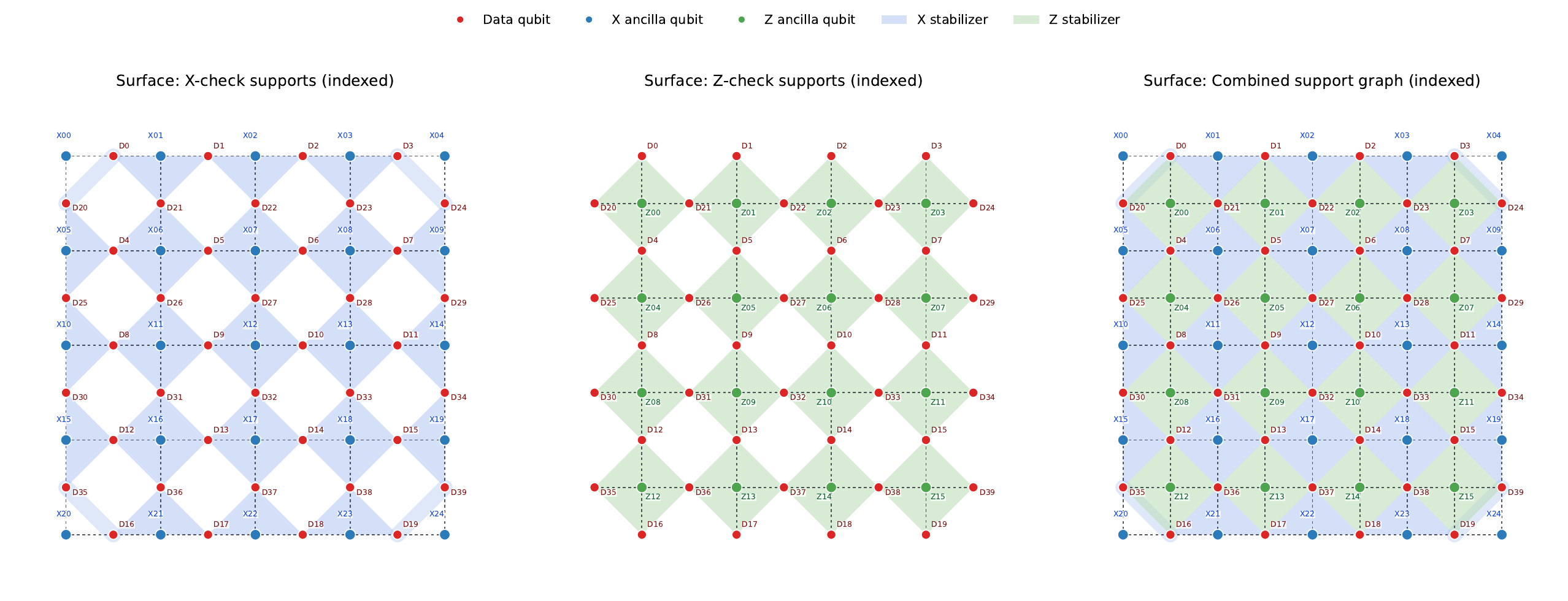}
        }
        \caption{Indexed outer-code support layouts used in request generation and replay analysis.}
        \label{fig:p4_surface_outer_layout}
    \end{figure*}

\begin{figure*}[t]
        \centering
        \subfloat[Digitized-GKP outer support topology with indexed data and check nodes.]{
            \includegraphics[width=0.48\textwidth]{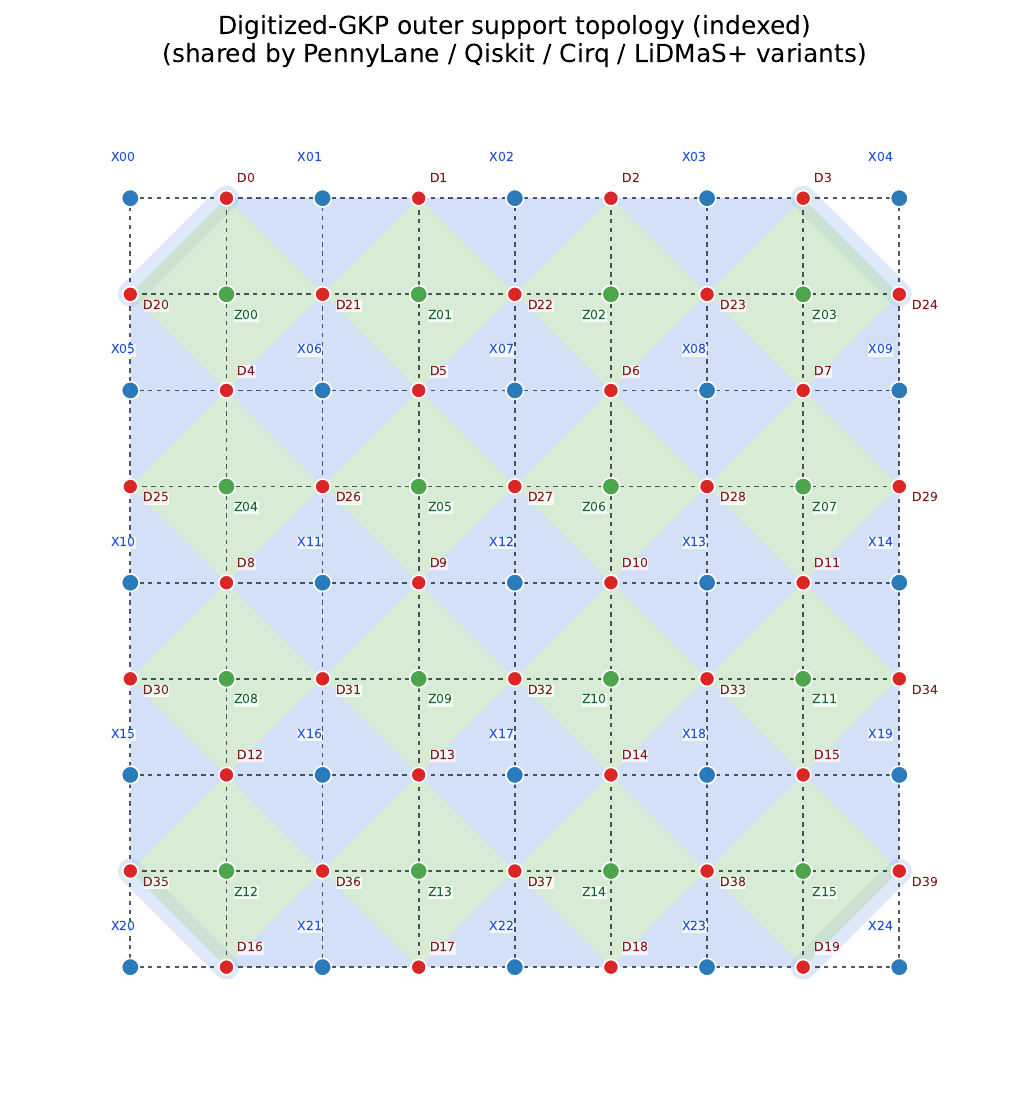}
        }
        \hfill
        \subfloat[Inner GKP code shown as an indexed single-mode phase-space lattice with decision boundaries.]{
            \includegraphics[width=0.48\textwidth]{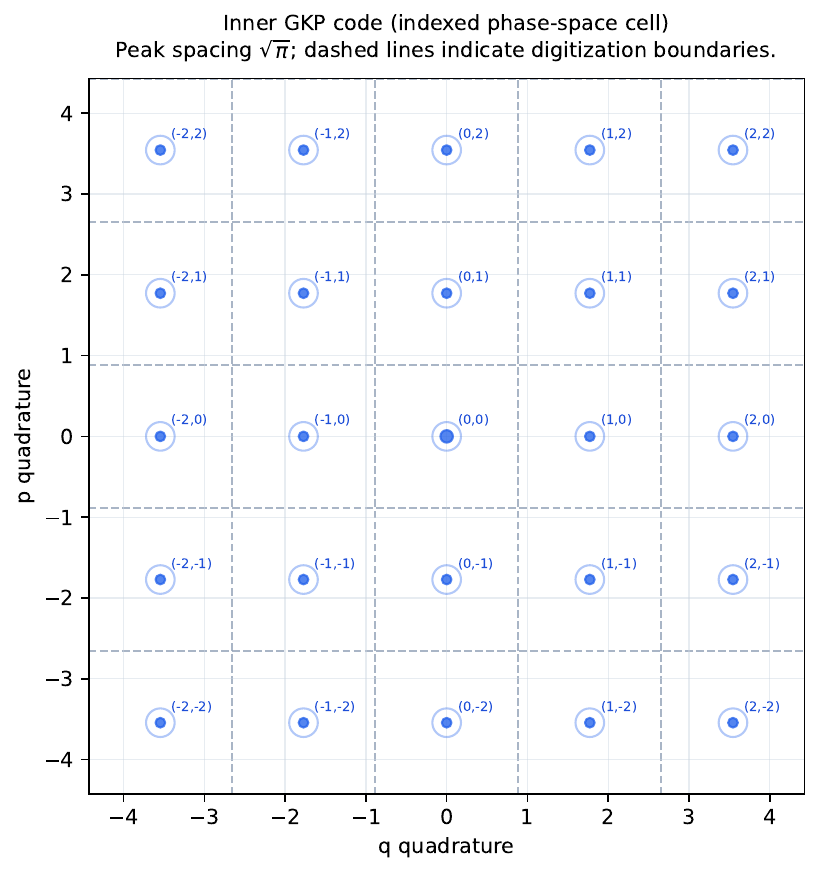}
        }
        \caption{GKP outer/inner structural views. The left panel shows the shared outer support graph; the right panel shows the conceptual inner-code phase-space representation.}
        \label{fig:p4_gkp_outer_inner_split}
    \end{figure*}

\begin{figure*}[t]
        \centering
        \includegraphics[width=0.95\textwidth]{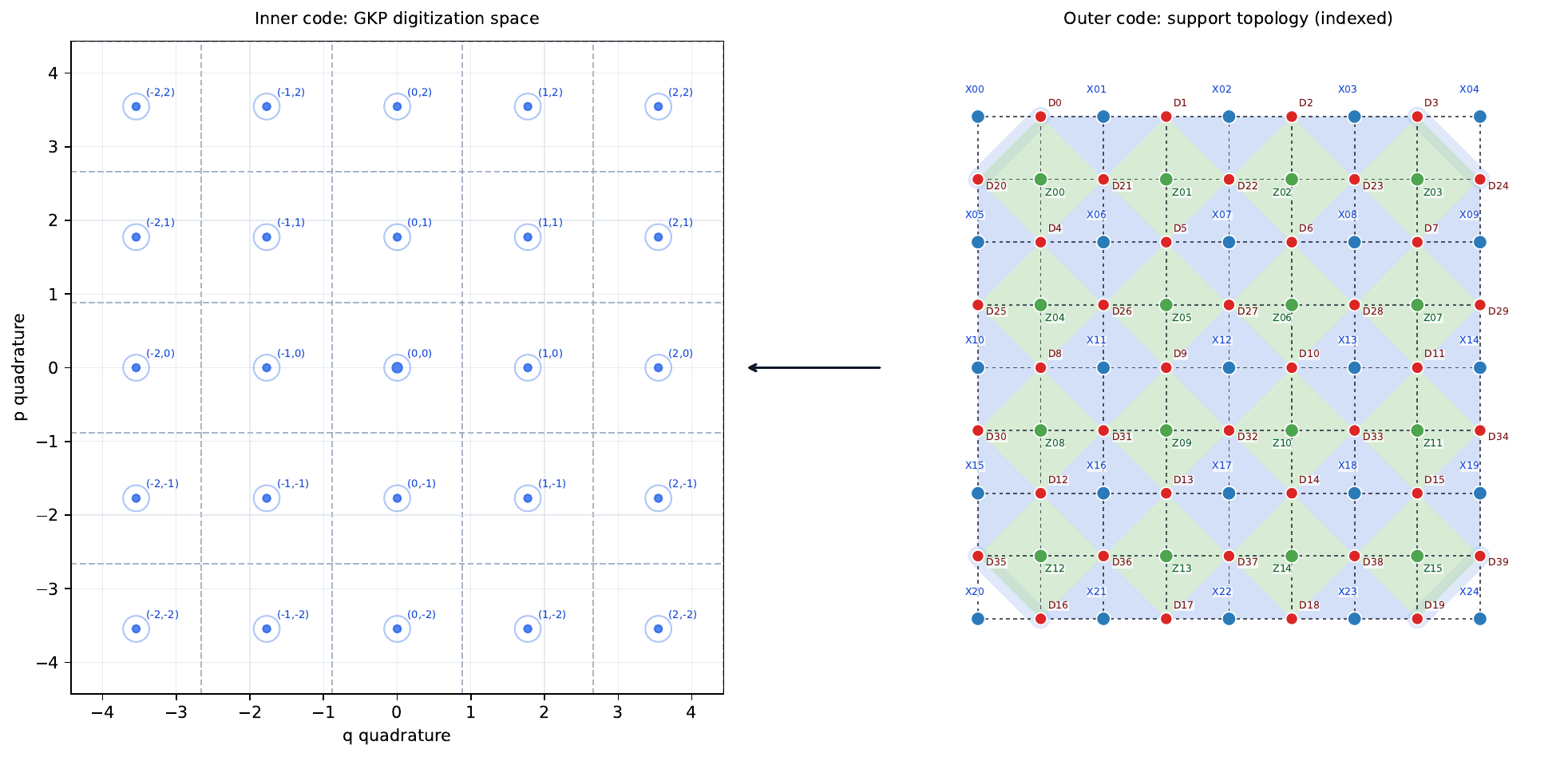}
        \caption{Concatenated schematic linking inner GKP digitization (left) to indexed outer syndrome-support topology (right).}
        \label{fig:p4_gkp_concatenated}
    \end{figure*}

    The following figures use the current run geometry (\(d=5\), rounds \(=4\)).
    They document the outer support graph and the GKP inner-code interpretation.
    Figures~\ref{fig:p4_surface_outer_layout},
    \ref{fig:p4_gkp_outer_inner_split}, and~\ref{fig:p4_gkp_concatenated} provide
    the indexed surface layout, the split GKP outer/inner view, and the
    concatenated inner-to-outer schematic.

    \clearpage
    \bibliographystyle{apsrev4-2}
    \bibliography{lidmas}

\end{document}